\documentclass[journal]{IEEEtran}

\ifCLASSINFOpdf
\else
    \usepackage[dvips]{graphicx}
\fi
\usepackage{url}
\usepackage{graphicx}

\usepackage{cite}
\usepackage[acronym]{glossaries}
\usepackage{siunitx}
\usepackage{orcidlink}
\usepackage{booktabs}
\usepackage{dsfont}
\usepackage{enumitem}
\usepackage{microtype}

\ifCLASSOPTIONcompsoc
  \usepackage[caption=false,font=normalsize,labelfont=sf,textfont=sf]{subfig}
\else
  \usepackage[caption=false,font=footnotesize]{subfig}
\fi

\DeclareSIUnit{\nothing}{\relax}

\DeclareSIUnit[quantity-product = ]\percent{\char`\%}

\makeatletter
\def\bstctlcite{\@ifnextchar[{\@bstctlcite}{\@bstctlcite[@auxout]}}
\def\@bstctlcite[#1]#2{\@bsphack
  \@for\@citeb:=#2\do{%
    \edef\@citeb{\expandafter\@firstofone\@citeb}%
    \if@filesw\immediate\write\csname #1\endcsname{\string\citation{\@citeb}}\fi}%
  \@esphack}
\makeatother

\newacronym{gpu}{GPU}{graphics processing unit}
\newacronym{2d}{2D}{two-dimensional}
\newacronym{asr}{ASR}{automatic speech recognition}
\newacronym{sid}{SID}{speaker identification}
\newacronym{nmt}{NMT}{neural machine translation}
\newacronym{tts}{TTS}{text-to-speech}
\newacronym{nlp}{NLP}{natural language processing}
\newacronym{dnn}{DNN}{deep neural network}
\newacronym{rnn}{RNN}{recurrent neural network}
\newacronym{cnn}{CNN}{convolutional neural network}
\newacronym{tcn}{TCN}{temporal convolutional network}
\newacronym{lstm}{LSTM}{long short-term memory}
\newacronym{ffnn}{FFNN}{feedforward neural network}
\newacronym{svm}{SVM}{support vector machine}
\newacronym{gan}{GAN}{generative adversarial network}
\newacronym{vae}{VAE}{variational autoencoder}
\newacronym{nf}{NF}{normalizing flow}
\newacronym{ema}{EMA}{exponential moving average}
\newacronym{relu}{ReLU}{rectified linear unit}
\newacronym{gmm}{GMM}{Gaussian mixture model}
\newacronym{mvdr}{MVDR}{minimum variance distortionless response}
\newacronym{mwf}{MWF}{multi-channel Wiener filter}
\newacronym{ic}{IC}{interaural coherence}
\newacronym{ild}{ILD}{interaural level difference}
\newacronym{itd}{ITD}{interaural time difference}
\newacronym{irm}{IRM}{ideal ratio mask}
\newacronym{cirm}{cIRM}{complex-valued ideal ratio mask}
\newacronym{ibm}{IBM}{ideal binary mask}
\newacronym{logfbe}{log-FBE}{log-filterbank energy}
\newacronym{mse}{MSE}{mean squared error}
\newacronym{mmse}{MMSE}{minimum mean squared error}
\newacronym{rms}{RMS}{root mean square}
\newacronym{snr}{SNR}{signal-to-noise ratio}
\newacronym{sisnr}{SI-SNR}{scale-invariant signal-to-noise ratio}
\newacronym{drr}{DRR}{direct-to-reverberant energy ratio}
\newacronym{pesq}{PESQ}{perceptual evaluation of speech quality}
\newacronym{stoi}{STOI}{short-term objective intelligibility}
\newacronym{estoi}{ESTOI}{extended short-term objective intelligibility}
\newacronym{tf}{T-F}{time-frequency}
\newacronym{stft}{STFT}{short-time Fourier transform}
\newacronym{brir}{BRIR}{binaural room impulse response}
\newacronym{zpr}{ZPR}{zero-padding rate}
\newacronym{sde}{SDE}{stochastic differential equation}
\newacronym{ve}{VE}{variance exploding}
\newacronym{vp}{VP}{variance preserving}
\newacronym{pc}{PC}{predictor-corrector}
\newacronym{iid}{IID}{independent and identically distributed}
\newacronym{ood}{OOD}{out-of-distribution}
\newacronym{dg}{DG}{domain generalization}
\newacronym{ou}{OU}{Ornstein-Uhlenbeck}
\newacronym{pdf}{PDF}{probability density function}
\newacronym{gln}{gLN}{global layer normalization}
\newacronym{cln}{cLN}{cumulative layer normalization}
\newacronym{elbo}{ELBO}{evidence lower bound}
\newacronym{gru}{GRU}{gated recurrent unit}

\newcommand*\dpesq{\Delta\text{\gls{pesq}}}
\newcommand*\destoi{\Delta\text{\gls{estoi}}}
\newcommand*\dsnr{\Delta\text{\gls{snr}}}
\newcommand*\inlineeq{\!=\!}
\newcommand*\inlineplus{\!+\!}
\newcommand*\inlinecdot{\!\cdot\!}

\newcommand*\inlinein{\!\in\!}
\newcommand*\verythinnegativespace{\mskip-0.5\thinmuskip}

\title{The Effect of Training Dataset Size on Discriminative and Diffusion-Based Speech Enhancement Systems}

\author{
Philippe~Gonzalez\,\orcidlink{0009-0006-4965-3514},
Zheng-Hua~Tan\,\orcidlink{0000-0001-6856-8928},
Jan~{\O}stergaard\,\orcidlink{0000-0002-3724-6114},
Jesper~Jensen\,\orcidlink{0000-0003-1478-622X},
Tommy~Sonne~Alstr{\o}m\,\orcidlink{0000-0003-0941-3146},
Tobias~May\,\orcidlink{0000-0002-5463-5509}
\thanks{Philippe Gonzalez and Tobias May are with the Department of Health Technology, Technical University of Denmark, 2800 Lyngby, Denmark (e-mail: \href{mailto:phigon@dtu.dk}{phigon@dtu.dk}; \href{mailto:tobmay@dtu.dk}{tobmay@dtu.dk}).}
\thanks{Zheng-Hua Tan, Jan {\O}stergaard and Jesper Jensen are with the Department of Electronic Systems, Aalborg University, 9220 Aalborg, Denmark (e-mail: \href{mailto:zt@es.aau.dk}{zt@es.aau.dk}; \href{mailto:jo@es.aau.dk}{jo@es.aau.dk}; \href{mailto:jje@es.aau.dk}{jje@es.aau.dk}).}
\thanks{Tommy Sonne Alstr{\o}m is with the Department of Applied Mathematics and Computer Science, Technical University of Denmark, 2800 Lyngby, Denmark (e-mail: \href{mailto:tsal@dtu.dk}{tsal@dtu.dk}).}
}

\IEEEaftertitletext{\vspace{-\baselineskip}}

\begin{document}

\robustify\bfseries

\bstctlcite{IEEEexample:BSTcontrol}

\maketitle

\begin{abstract}
The performance of deep neural network-based speech enhancement systems typically increases with the training dataset size.
However, studies that investigated the effect of training dataset size on speech enhancement performance did not consider recent approaches, such as diffusion-based generative models.
Diffusion models are typically trained with massive datasets for image generation tasks, but whether this is also required for speech enhancement is unknown.
Moreover, studies that investigated the effect of training dataset size did not control for the data diversity.
It is thus unclear whether the performance improvement was due to the increased dataset size or diversity.
Therefore, we systematically investigate the effect of training dataset size on the performance of popular state-of-the-art discriminative and diffusion-based speech enhancement systems in matched conditions.
We control for the data diversity by using a fixed set of speech utterances, noise segments and binaural room impulse responses to generate datasets of different sizes.
We find that the diffusion-based systems perform the best relative to the discriminative systems in terms of objective metrics with datasets of \SI{10}{\hour} or less.
However, their objective metrics performance does not improve when increasing the training dataset size as much as the discriminative systems, and they are outperformed by the discriminative systems with datasets of \SI{100}{\hour} or more.
\end{abstract}


\begin{tikzpicture}[remember picture,overlay]
\node[anchor=south,yshift=3pt] at (current page.south) {
  \fbox{\parbox{\dimexpr\textwidth - 2\fboxsep}{
    \footnotesize \copyright 2023 IEEE. Personal use of this material is permitted. Permission from IEEE must be obtained for all other uses, in any current or future media, including reprinting/republishing this material for advertising or promotional purposes, creating new collective works, for resale or redistribution to servers or lists, or reuse of any copyrighted component of this work in other works.
  }}
};
\end{tikzpicture}
\vspace{-\baselineskip}

\IEEEpeerreviewmaketitle

\section{Introduction}

\IEEEPARstart{U}{nderstanding} speech in noisy and reverberant environments can be challenging for both normal-hearing and hearing-impaired listeners~\cite{nabelek1974monaural,nabelek1981effect}.
Therefore, speech enhancement, which aims to improve the intelligibility and quality of speech signals corrupted by noise and reverberation, is an integral part of many technical applications, such as hearing aids and communication systems.
The majority of newly-proposed speech enhancement systems are based on \glspl{dnn} due to their superior performance over traditional approaches~\cite{xu2015regression,wang2018supervised}.
These systems are commonly trained in a supervised manner with a large number of noisy and clean speech signals.
As the number of trainable parameters increases, \glspl{dnn} have the potential to capture more details in the probability distribution of the training data, but this requires training them with larger datasets.
For example, diffusion models~\cite{sohl2015deep,song2021score,ho2020denoising}, which have been recently applied to speech enhancement~\cite{lu2021study,lu2022conditional,welker2022speech,richter2023speech,yen2023cold,chen2023metric,qiu2023srtnet,tai2023dose,shi2024diffusion,gonzalez2023investigating,gonzalez2024diffusion,nortier2024unsupervised,ayilo2024diffusion}, are typically trained with huge datasets in image generation literature~\cite{wang2023patch}.
However, whether this is also required for speech enhancement is unknown.

Few studies have investigated the effect of training dataset size on the performance of state-of-the-art speech enhancement systems in a systematic way.
In~\cite{wang2013towards}, a \gls{ffnn} was trained with an increasing number of noises and utterances, which improved the classification performance of individual \glsentrylong{tf} units into speech and noise.
However, the dataset size increased with the number of noises and utterances used to generate the mixtures.
As a consequence, the effects of the size and diversity of the training data were entangled.
In~\cite{xu2015regression}, a \gls{ffnn} was trained with datasets of different sizes generated from a fixed number of noises and utterances, i.e.\ fixed diversity.
Speech quality results increased with the training dataset size, but since the utterances were selected from TIMIT~\cite{garofolo1993timit}, whose training split is only \SI{4}{\hour}-long, the performance saturated for datasets larger than \SI{100}{\hour} due to the increased redundancy in the training data.
In~\cite{chen2016large}, a \gls{ffnn} was trained with a fixed number of mixtures, i.e.\ fixed dataset size, using either \SI{100}{\nothing} or \SI{10000}{\nothing} noises.
The system strongly benefited from the increased number of noises and matched the performance of a noise-specific system, but the effect of training dataset size was not investigated.
In~\cite{chai2021cross}, a bidirectional \gls{gru} network was trained with datasets of different sizes generated using an increasing range of \glspl{snr}, and the performance of a downstream speech recognition system improved consequently.
However, by increasing the range of \glspl{snr} seen during training, the acoustic mismatch between training and testing was reduced.
The performance improvement was thus attributed to the reduced mismatch rather than the increased dataset size.
In~\cite{rehr2021snr}, a \gls{ffnn} and a recurrent \gls{lstm} network were trained with \SI{100}{\hour}-long datasets generated from different noise databases.
The systems performed the best when trained with datasets generated from large and diverse noise databases.
While the study differentiated between the size and diversity of the data, it referred to the size of the noise databases used to generate the mixtures, rather than the amount of mixtures generated to train the systems, which was kept constant.
In summary, most studies have not investigated the effect of the training dataset size independently of its diversity, and have considered outdated \glspl{ffnn} or recurrent neural networks.

In this study, we systematically investigate the effect of training dataset size on the performance of popular state-of-the-art speech enhancement systems in matched conditions.
We consider three discriminative systems, namely Conv-TasNet~\cite{luo2019conv}, DCCRN~\cite{hu2020dccrn} and MANNER~\cite{park2022manner}, and three diffusion-based approaches, namely SGMSE+~\cite{richter2023speech}, SGMSE+M~\cite{lemercier2023analysing} and the system from~\cite{gonzalez2023investigating,gonzalez2024diffusion}.
To control for the data diversity, we generate mixture datasets of different sizes using a fixed set of speech utterances, noise segments and \glspl{brir}.
This way, the training distribution is fixed, and only the number of training examples is changed.
Code and audio examples are available online\footnote{\url{https://github.com/philgzl/brever} and \url{https://philgzl.github.io/lst}}.

\section{Signal model}

Let ${s}$ denote a clean speech signal and ${\{n_i\}_{i=1}^{N}}$ a set of noise signals where ${N}$ is the number of noise sources in the acoustic scene.
The mixture ${x_L}$ at the left ear of a binaural receiver in the acoustic scene can be expressed as follows,
\begin{align}
    x_L &= s \ast h_{s,L} + \sum_{i=1}^{N} n_i \ast h_{n_i,L},
\end{align}
where ${h_{s,L}}$ is the left channel of the \gls{brir} between the receiver and the speech source, and ${h_{n_i,L}}$ is the left channel of the \gls{brir} between the receiver and the ${i}$-th noise source.
Defining the target signal for the speech enhancement system requires choosing a reflection boundary ${b}$ beyond which speech reflections are considered detrimental to speech intelligibility~\cite{bradley2003importance,roman2013speech}.
Let ${\mathds{1}_{[a, b[}}$ denote the indicator function of a time interval ${[a, b[}$.
By denoting ${h_{s,L}^{\mathrm{early}} \inlineeq h_{s,L} \inlinecdot \mathds{1}_{[0, b[}}$ the part of ${h_{s,L}}$ up to time instant ${b}$ and ${h_{s,L}^{\mathrm{late}} \inlineeq h_{s,L} \inlinecdot \mathds{1}_{[b, \infty[}}$ the part of ${h_{s,L}}$ beyond time instant ${b}$, the target signal ${y_L}$ for the speech enhancement system at the left ear is
\begin{equation}
    y_L = s \ast h_{s,L}^{\mathrm{early}},
\end{equation}
while the interfering signal ${n_L}$ at the left ear includes the late speech reflections and the reverberant noise,
\begin{equation}
    n_L = s \ast h_{s,L}^{\mathrm{late}} + \sum_{i=1}^{N} n_i \ast h_{n_i,L}.
\end{equation}
Note that since ${h_{s,L} \inlineeq {h_{s,L}^{\mathrm{early}} \inlineplus h_{s,L}^{\mathrm{late}}}}$, we have ${x_L \inlineeq y_L \inlineplus n_L}$ due to the linearity of the convolution operator.
Similar expressions are obtained for the right ear signals ${x_R}$, ${y_R}$ and ${n_R}$.

As the speech enhancement systems considered in this study are single-channel, we average the left and right channels of the different signals.
We denote the single-channel mixture, target and interfering signals as ${x \inlineeq \frac{x_L + x_R}{2}}$, ${y \inlineeq \frac{y_L + y_R}{2}}$ and ${n \inlineeq \frac{n_L + n_R}{2}}$ respectively.
Even though the channels are averaged, a binaural model allows to simulate realistic mixtures where the position of the sources contributes to the acoustic diversity.

\section{Dataset generation}

We generate noisy and reverberant mixtures at \SI{16}{\kilo\hertz} using speech utterances, noise segments and \glspl{brir} from multiple corpora and databases.
The speech utterances are selected from TIMIT~\cite{garofolo1993timit}, LibriSpeech (100-hour version)~\cite{panayotov2015librispeech}, WSJ SI-84~\cite{paul1992design}, Clarity~\cite{cox2022clarity} and VCTK~\cite{veaux2013voice}.
The noises are selected from TAU~\cite{heittola2019tau}, NOISEX~\cite{varga1993noisex}, ICRA~\cite{dreschler2001icra}, DEMAND~\cite{thiemann2013demand} and ARTE~\cite{weisser2019ambisonic}.
The \glspl{brir} are selected from Surrey~\cite{hummersone2010surrey}, ASH~\cite{shanon2021ash}, BRAS~\cite{brinkmann2021bras}, CATT~\cite{catt_brirs} and AVIL~\cite{marschall2017database}.
Details about each corpus and database are provided in Tab.~\ref{tab:corpora}.
Each acoustic scene is simulated by placing one speech source and ${N \inlinein \{1, \verythinnegativespace 2, \verythinnegativespace 3\}}$ noise sources at random spatial locations in the same room, uniformly distributed in [\SI{-90}{\degree},\,\SI{90}{\degree}] in front of the receiver.
The reflection boundary is set to ${b \inlineeq \SI{50}{\milli\second}}$ to include early reflections in the target signal ${y}$ as suggested in~\cite{bradley2003importance,roman2013speech}.
The target and interfering signals ${y}$ and ${n}$ are mixed at a random \gls{snr} uniformly distributed in [\SI{-5}{\decibel},\,\SI{10}{\decibel}].

\begin{table}
    \centering
    \caption{Corpora and databases details}
    \label{tab:corpora}
    \vspace{-\baselineskip}
    \subfloat[Speech corpora]{
        \setlength{\tabcolsep}{3.5pt}
        \begin{tabular}{
            c
            S[table-format=3]
            S[table-format=5,group-minimum-digits=4]
            S[table-format=3.1]
            S[table-format=2.1]
            S[table-format=1.1]
            S[table-format=1.1]
        }
            \toprule
            Corpus & {Speakers} & {Utterances} & {Hours} & {Avg.\,len.} & {Min.\,len.} & {Max.\,len.}\\
            \midrule
            TIMIT   & 630 &  6300 &   5.4 &  3.1\,s & 0.9\,s &  7.8\,s \\
            Libri.  & 251 & 28539 & 100.6 & 12.7\,s & 1.4\,s & 24.5\,s \\
            WSJ     & 131 & 34738 &  69.5 &  7.2\,s & 0.9\,s & 44.8\,s \\
            Clarity &  40 & 11352 &   8.9 &  2.8\,s & 1.2\,s &  7.7\,s \\
            VCTK    & 110 & 44455 &  41.6 &  3.4\,s & 1.2\,s & 16.6\,s \\
            \bottomrule
        \end{tabular}
    }\hfil
    \subfloat[Noise databases]{
        \begin{tabular}{
            c
            S[table-format=2]
            S[table-format=2.1]
        }
            \toprule
            Database & {Types} & {Hours} \\
            \midrule
            TAU    & 10 & 40.0 \\
            NOISEX & 15 &  1.0 \\
            ICRA   & 10 &  1.1 \\
            DEMAND & 18 &  1.5 \\
            ARTE   & 13 &  0.5 \\
            \bottomrule
        \end{tabular}
    }\hfil
    \subfloat[\gls{brir} databases]{
        \begin{tabular}{
            c
            S[table-format=2]
            S[table-format=3]
        }
            \toprule
            Database & {Rooms} & {\glspl{brir}} \\
            \midrule
            Surrey &  4 & 148 \\
            ASH    & 35 & 538 \\
            BRAS   &  4 & 180 \\
            CATT   & 11 & 407 \\
            AVIL   &  4 &  96 \\
            \bottomrule
        \end{tabular}
    }
    \vspace{-2pt}
\end{table}

To generate a training dataset, we randomly select speech utterances, noise segments and \glspl{brir} from a subset of each speech corpus, noise database and \gls{brir} database respectively.
For the speech, this subset is constructed by randomly selecting \SI{80}{\percent} of the utterances from each corpus.
For the noise, we sample segments within \SI{80}{\percent} of the length of each file.
For the \glspl{brir}, we select every other \gls{brir} in each room.
The validation dataset uses the same subset of speech utterances, noise segments and \glspl{brir} as the training dataset, but consists of different random mixture realizations.
The test dataset is generated using the remaining set of speech utterances, noise segments and \glspl{brir}.
Note that utterances from the same speaker, segments from the same noise file and \glspl{brir} from the same room can be used for training and testing.
This is deliberate, as we are interested in evaluating the system in matched conditions, such that distribution shifts between training and testing are minimized and only the effect of the number of training examples is captured.
We generate five training datasets with sizes of \SI{3}{\hour}, \SI{10}{\hour}, \SI{30}{\hour}, \SI{100}{\hour} and \SI{300}{\hour} respectively.
The number of mixtures in each dataset is \SI{2595}{\nothing}, \SI{8660}{\nothing}, \SI{25930}{\nothing}, \SI{86074}{\nothing} and \SI{258259}{\nothing} respectively.
The validation and test datasets are both fixed to \SI{30}{\minute}.
They consist of 305 and 310 mixtures respectively.

When randomly selecting a corpus from which to pick a utterance to form a mixture, one option is to use equal probabilities for each corpus.
While this would result in a similar number of utterances from each corpus, it would result in a large imbalance in terms of duration, since the five considered speech corpora have very different utterance length distributions as shown in Tab.~\ref{tab:corpora}.
To avoid this, we weight the probability of selecting a corpus by the inverse of its average utterance length.
This is only done for the training dataset, as the systems are evaluated on the validation and test datasets on a per-mixture basis.
This is not an issue when selecting noise segments and \glspl{brir}, since the mixture length is defined by the speech utterance length.
We thus use equal probabilities for selecting the noise and \gls{brir} databases.

The speech utterances, noise segments and \glspl{brir} are randomly drawn with replacement.
This means that the same speech utterance can be used to generate multiple mixtures in the same dataset, and the number of repetitions increases with the dataset size.
Moreover, as the number of utterances in each corpus varies substantially, utterances are more or less likely to be repeated depending on which corpus they are selected from.
Note that the mixtures are still unique, as the chances of selecting the same speech utterance, noise segment, \gls{brir}, \gls{snr} and spatial locations twice are very low.
Table~\ref{tab:repetitions} shows the percentage of mixtures using repeated speech utterances in each training dataset, and the corpus they are selected from.
A similar analysis can be made for the noise segments and the \glspl{brir}, but is not shown here for brevity.

\begin{table}
    \centering
    \caption{Percentage of mixtures using repeated speech utterances in each training dataset in terms of duration}
    \label{tab:repetitions}
    \begin{tabular}{S[table-format=3]*{6}{S[table-format=2]}}
        \toprule
        {Dataset} & {TIMIT} & {WSJ} & {Clarity} & {Libri.} & {VCTK} & {Total} \\
        \midrule
          3\,h &  2\% &  1\% &  2\% &  0\% & 0\% &  5\% \\
         10\,h &  8\% &  1\% &  5\% &  0\% & 1\% & 15\% \\
         30\,h & 15\% &  3\% & 12\% &  1\% & 3\% & 34\% \\
        100\,h & 20\% &  9\% & 19\% &  5\% & 9\% & 61\% \\
        300\,h & 20\% & 15\% & 20\% & 11\% & 17\% & 83\% \\
        \bottomrule
    \end{tabular}
\end{table}

\section{Systems, training and objective metrics}

We evaluate three different discriminative speech enhancement systems, namely Conv-TasNet~\cite{luo2019conv}, DCCRN~\cite{hu2020dccrn} and MANNER~\cite{park2022manner}.
They have \SI{4.9}{\mega\nothing}, \SI{3.7}{\mega\nothing} and \SI{21.2}{\mega\nothing} parameters respectively.
We also evaluate three diffusion-based systems, namely SGMSE+~\cite{richter2023speech}, SGMSE+M~\cite{lemercier2023analysing} and the system from~\cite{gonzalez2023investigating,gonzalez2024diffusion}, which we denote as SGMSE+M$_{\text{Heun}}^{\text{cos}}$.
These three systems all use the NCSN++ architecture from~\cite{song2021score} for the score network.
SGMSE+M and SGMSE+M$_{\text{Heun}}^{\text{cos}}$ use a smaller version of NCSN++ (\SI{27.8}{\mega\nothing} parameters) compared to SGMSE+ (\SI{66.1}{\mega\nothing} parameters), as this was reported to reduce the computational cost without degrading performance~\cite{lemercier2023analysing}.
Compared to SGMSE+M, SGMSE+M$_{\text{Heun}}^{\text{cos}}$ uses a cosine noise schedule, a Heun-based sampler and a different preconditioning~\cite{gonzalez2023investigating,gonzalez2024diffusion}.
The number of sampling steps is fixed to 64.
Technical details about the implementation of both discriminative and diffusion-based systems can be found in~\cite{gonzalez2023investigating}.

The systems are trained with the different datasets for the same number of neural network parameter updates.
That is, as the dataset size increases, the number of training epochs is proportionally reduced.
This allows for a fair comparison, since the systems are trained for the same amount of time.
The number of epochs is set to 1000 for the \SI{3}{\hour} dataset, 300 for the \SI{10}{\hour} dataset, 100 for the \SI{30}{\hour} dataset, 30 for the \SI{100}{\hour} dataset and 10 for the \SI{300}{\hour} dataset.
The experiment is repeated three times with different random neural network parameter initializations, and the metrics are averaged across repetitions.

The systems are evaluated in terms of \gls{pesq}~\cite{recommendation2001perceptual}, \gls{estoi}~\cite{jensen2016algorithm} and \gls{snr}.
The results are reported in terms of average objective metric improvement from the input mixture to the enhanced output.
The improvements are denoted as ${\dpesq}$, ${\destoi}$ and ${\dsnr}$ respectively.

\section{Results}

\begin{figure*}
    \centering
    \includegraphics{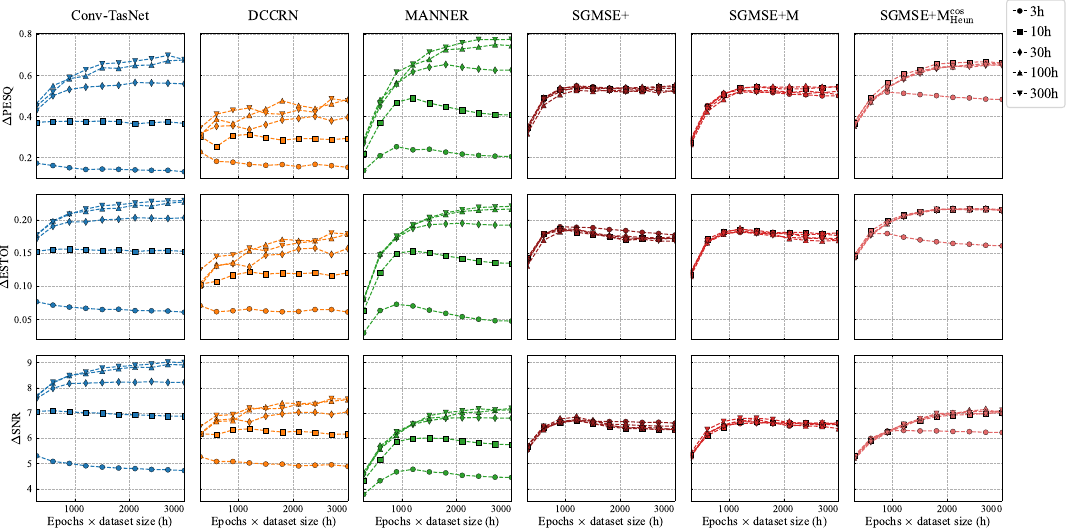}%
    \caption{${\dpesq}$, ${\destoi}$ and ${\dsnr}$ results on the test dataset as a function of the number of epochs ${\times}$ the training dataset size in hours.}
    \label{fig:training_curves}
\end{figure*}

\begin{figure}
    \centering
    \includegraphics{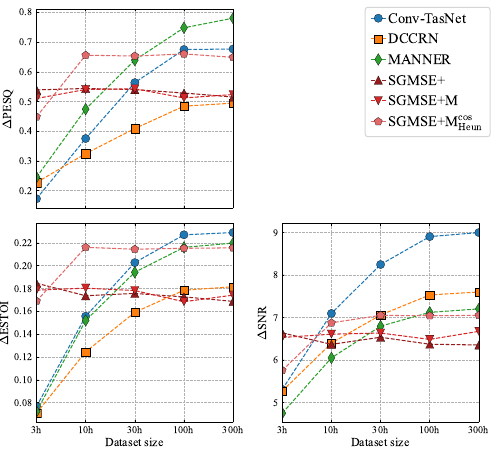}%
    \caption{${\dpesq}$, ${\destoi}$ and ${\dsnr}$ results as a function of the training dataset size.}
    \label{fig:summary}
\end{figure}

Figure~\ref{fig:training_curves} shows the performance on the test dataset as training progresses for each system and each dataset size.
It can be seen that the discriminative systems, i.e.\ Conv-TasNet, DCCRN and MANNER, strongly benefit from increasing the training dataset size, despite the increasing amount of utterance repetitions reported in Tab.~\ref{tab:repetitions}.
More specifically, these systems heavily overfit when training with the \SI{3}{\hour} dataset for too many epochs.
As the training dataset size increases, they stop overfitting and benefit from longer training.
Performance improvements with the dataset size can be observed until \SI{300}{\hour}.
Meanwhile, the diffusion-based systems, i.e.\ SGMSE+, SGMSE+M and SGMSE+M$_{\text{Heun}}^{\text{cos}}$, show a very different behavior.
With the \SI{3}{\hour} dataset, they overfit less severely and show substantially higher results compared to the discriminative systems.
While the performance of SGMSE+M$_{\text{Heun}}^{\text{cos}}$ improves when increasing the dataset size to \SI{10}{\hour}, they show very similar curves when further increasing the dataset size beyond \SI{10}{\hour}. 
Despite having more parameters, SGMSE+ shows similar performance to SGMSE+M, which is in line with~\cite{lemercier2023analysing}.

The best model for each training run is selected based on the validation loss.
The performance of the selected model on the test dataset is plotted as a function of the training dataset size in Figure~\ref{fig:summary}.
Similar to the previous results, the performance of the discriminative systems substantially improves with the dataset size.
However, while the diffusion-based systems outperform the discriminative systems with the \SI{3}{\hour} and \SI{10}{\hour} datasets, they do not improve when further increasing the dataset size.
With the \SI{100}{\hour} and \SI{300}{\hour} datasets, they are outperformed by Conv-TasNet and MANNER in terms of ${\dpesq}$, and by Conv-TasNet and DCCRN in terms of ${\dsnr}$.
Since SGMSE+ shows the same behavior as SGMSE+M, these results cannot be explained by a lack of model capacity.

\section{Discussion}

The results suggest that diffusion-based speech enhancement systems are remarkably suited when a small amount of training data is available, as SGMSE+, SGMSE+M and SGMSE+M$_{\text{Heun}}^{\text{cos}}$ perform the best relative to the discriminative systems when trained with the \SI{3}{\hour} and \SI{10}{\hour} datasets.
Our results are in contrast to image generation literature, where diffusion models are typically trained with datasets of billions of images~\cite{wang2023patch}.
We hypothesize on possible explanations for our results:
\begin{itemize}[leftmargin=*]
    \item The speech enhancement task is very different from the image generation task.
    In image generation, the model is tasked with generating coherent images from scratch given a text prompt, and multiple valid yet very different images can be generated from the same prompt.
    In speech enhancement, the model is provided with a mixture that has the same modality and dimensionality as the clean speech, and can thus leverage a lot of information from the input to generate the output.
    While multiple versions of the clean speech can be generated from the same mixture, these versions should not be very different from each other, as the output speech should be coherent with the mixture.
    \item The stochastic nature of the diffusion process acts as a strong regularizer, which allows the model to perform well despite being trained with a small amount of data.
    Indeed, the neural network is presented with training examples mixed with random Gaussian noise realizations at different levels during training.
    This is in line with \cite{kingma2023understanding}, which showed that the diffusion model objective is equivalent to the \gls{elbo} with data augmentation consisting of Gaussian noise perturbation.
    This also explains the similar curves in Fig.~\ref{fig:training_curves} for the \SI{10}{\hour}, \SI{30}{\hour}, \SI{100}{\hour} and \SI{300}{\hour} datasets.
    \item While SGMSE+M$_{\text{Heun}}^{\text{cos}}$ shows superior performance compared to SGMSE+M thanks to the updated noise schedule, sampler and preconditioning, there might be other design aspects that prevent the system from scaling. E.g.\ the NCSN++ architecture, which was borrowed from image generation literature, might not be optimal for speech processing.
\end{itemize}

\section{Conclusion}

We investigated the effect of training dataset size on the performance of three discriminative and three diffusion-based speech enhancement systems in matched conditions.
We found that the diffusion-based systems performed the best relative to the discriminative systems in terms of objective metrics with datasets of \SI{10}{\hour} or less, but they were outperformed by the discriminative systems with datasets of \SI{100}{\hour} or more.
This suggests that diffusion-based approaches are remarkably suited when a small amount of training data is available.
However, this also suggests that they do not benefit from increasing the training dataset size as much as discriminative systems.
Future work should investigate if the conclusions generalize to unseen speaker, noise and room conditions.
In addition, a formal listening test should be conducted to investigate if the reported differences in terms of objective metrics are perceptually relevant.

\bibliographystyle{IEEEtran}
\bibliography{bib/IEEEabrv, bib/abbrv, bib/refs}
\end{document}